\documentstyle [12pt]{article}
\makeatletter \oddsidemargin  0in \evensidemargin 0in
\textwidth16cm \RequirePackage[dvips]{graphicx} \textheight 20.5cm
\newcommand{\singlespacing}{\let\CS=\@currsize\renewcommand{\baselinestretch}{1.5}\tiny\CS}
\makeatletter \oddsidemargin  -.1in \evensidemargin -.1in
\newcommand{\doublespacing}{\let\CS=\@currsize\renewcommand{\baselinestretch}{1.35}\tiny\CS}

\def\@citex[#1]#2{\if@filesw\immediate\write\@auxout{\string\citation{#2}}\fi
  \def\@citea{}\@cite{\@for\@citeb:=#2\do
    {\@citea\def\@citea{,\linebreak[0]\hskip0pt plus .2em}%
      \@ifundefined{b@\@citeb}%
    {{\bf ?}\@warning{Citation `\@citeb' on page \thepage\space undefined}}%
      \hbox{\csname b@\@citeb\endcsname}}}{#1}}

\newtheorem{rule-def}[theorem]{Rule}
\begin{document}
\title{\bf Impossibility of partial swapping of Quantum
Information}\author{I.Chakrabarty\thanks{Corresponding author:
E-Mail-indranilc@indiainfo.com }\\ Department of
Mathematics\\Heritage Institute of Technology,Kolkata-107,West
Bengal,India}
\date{}
\maketitle{}
\begin{abstract} It is a well known fact that a quantum  state
$|\psi(\theta,\phi)\rangle$ is represented by a point on the Bloch
sphere, characterized by two parameters $\theta$ and $\phi$. Here
in this work, we find out another impossible operation in quantum
information theory . We name this impossibility as 'Impossibility
of partial swapping of quantum information '. By this we mean that
if two unknown quantum states are given at the input port, there
exists no physical process , consistent with the principles of
quantum mechanics, by which we can partially swap either of the
two parameters $\theta$ and $\phi$ between these two quantum
states. In this work we provided the impossibility proofs for the
qubits( i.e the quantum states taken from two dimensional Hilbert
space) and this impossible operation can be shown to hold in
higher dimension also.
\end{abstract}
\section{Introduction:} In quantum information theory it is most important to know
various differences between the classical and quantum information.
Quantum information theory on one hand broadens the set consisting
of information processing protocols, but also on other hand puts
restrictions  on various operations which are feasible in
classical information theory [1-3]. These restrictions on many
quantum operations is making quantum information more secure. In
the famous land mark paper of Wootters and Zurek it was shown that
a single quantum cannot be cloned [1]. Later it was also shown by
Pati and Braunstein that we cannot delete either of the two
quantum states when we are provided with two identical quantum
states at our input port [2]. In spite of these two famous
'no-cloning' [1] and 'no-deletion' [2] theorem there are many
other 'no-go' theorems like 'no-self replication' [3] ,
'no-partial erasure' [4], 'no-splitting' [5] and many more which
have come up.\\
In this work we claim another impossibility in quantum information
theory. This 'no-go' theorem is yet another addition to the set of
forbidden operations in quantum information theory. We know that
the quantum state $|\psi(\theta,\phi)\rangle$ can be represented
as a point on the Bloch-sphere, characterized by two parameters
$\theta$ and $\phi$. Here in this work we define a concept called
partial swapping of quantum information, which is not the same as
the swapping of two quantum states. As the information content in
a qubit is dependent on the angles $\theta$ and $\phi$, the
partial swapping of quantum parameters $\theta$ and $\phi$ is
given by,
\begin{eqnarray}
|A(\theta_1,\phi_1)\rangle|\bar{A}(\bar{\theta_1},\bar{\phi_1})\rangle\longrightarrow
|A(\theta_1,\bar{\phi_1})\rangle|\bar{A}(\bar{\theta_1},\phi_1)\rangle\\
|A(\theta_1,\phi_1)\rangle|\bar{A}(\bar{\theta_1},\bar{\phi_1})\rangle\longrightarrow
|A(\bar{\theta_1},\phi_1)\rangle|\bar{A}(\theta_1,\bar{\phi_1})\rangle
\end{eqnarray}
Here in this work we will show that the above transformations will
not be possible in quantum regime. The linear structure of quantum
mechanics prohibits the execution of these operations. In the next
section we will provide the impossibility proof of partial
swapping of quantum parameters $\{\theta,\phi\}$ for two
dimensional quantum states (i.e qubits) and also discussed about
what will be the probable extension of the proof for quantum
states taken from a general 'd' dimensional Hilbert space. Then
the conclusion follows.
\section{Partial swapping of quantum information: Linearity of quantum mechanics}
Let us a consider  unknown qubits
$\{|X(\theta,\phi)\rangle,|\bar{X}(\bar{\theta},\bar{\phi})\rangle\}\in
H$, which can be expressed as a linear combination of basis
vectors $\{
|A(\theta_1,\phi_1)\rangle,|\bar{A}(\bar{\theta_1},\bar{\phi_1})\rangle\}$
of the two dimensional Hilbert space $H$.
\begin{eqnarray}
|X(\theta,\phi)\rangle=a|A(\theta_1,\phi_1)\rangle+b|\bar{A}(\bar{\theta_1},\bar{\phi_1})\rangle\nonumber\\
|\bar{X}(\bar{\theta},\bar{\phi})\rangle=c|A(\theta_1,\phi_1)\rangle+d|\bar{A}(\bar{\theta_1},\bar{\phi_1})\rangle
\end{eqnarray}
where,
\begin{eqnarray}
a^2+|b|^2=1\nonumber\\
c^2+|d|^2=1
\end{eqnarray}
(where $a^2,b^2$ and $c^2,d^2$ are  corresponding probabilities of
the qubits
$|X(\theta,\phi)\rangle,|\bar{X}(\bar{\theta},\bar{\phi})\rangle$
to collapse into the states $\{
|A(\theta_1,\phi_1)\rangle,|\bar{A}(\bar{\theta_1},\bar{\phi_1})\rangle
\}$, when it undergoes a measurement in the same basis).
 First of all let us
assume that  swapping of phase angles $\{\phi_1,\bar{\phi_1}\}$ be
possible for the orthogonal basis vectors $\{
|A(\theta_1,\phi_1)\rangle,|\bar{A}(\bar{\theta_1},\bar{\phi_1})\rangle\}$.
\begin{eqnarray}
|A(\theta_1,\phi_1)\rangle|A(\theta_1,\phi_1)\rangle\longrightarrow|A(\theta_1,\phi_1)\rangle|A(\theta_1,\phi_1)\rangle\nonumber\\
|A(\theta_1,\phi_1)\rangle|\bar{A}(\bar{\theta_1},\bar{\phi_1})\rangle\longrightarrow|A(\theta_1,\bar{\phi_1})\rangle|\bar{A}(\bar{\theta_1},\phi_1)\rangle\nonumber\\
|\bar{A}(\bar{\theta_1},\bar{\phi_1})\rangle|A(\theta_1,\phi_1)\rangle\longrightarrow|\bar{A}(\bar{\theta_1},\phi_1)\rangle|A(\theta_1,\bar{\phi_1})\rangle\nonumber\\
|\bar{A}(\bar{\theta_1},\bar{\phi_1})\rangle|\bar{A}(\bar{\theta_1},\bar{\phi_1})\rangle\longrightarrow|\bar{A}(\bar{\theta_1},\bar{\phi_1})\rangle|\bar{A}(\bar{\theta_1},\bar{\phi_1})\rangle
\end{eqnarray}
Then the action of the above transformations defined for the
orthogonal basis vectors, on the combined system of qubits
$|X(\theta,\phi)\rangle, |\bar{X}(\bar{\theta},\bar{\phi})\rangle$, is given by
\begin{eqnarray}
&&|X(\theta,\phi)\rangle
|\bar{X}(\bar{\theta},\bar{\phi})\rangle={}\nonumber\\&&\{a|A(\theta_1,\phi_1)\rangle+b|\bar{A}(\bar{\theta_1},\bar{\phi_1})\rangle\}
\{
c|A(\theta_1,\phi_1)\rangle+d|\bar{A}(\bar{\theta_1},\bar{\phi_1})\rangle\}={}\nonumber\\&&
ac|A(\theta_1,\phi_1)\rangle|A(\theta_1,\phi_1)\rangle+ad|A(\theta_1,\phi_1)\rangle|\bar{A}(\bar{\theta_1},\bar{\phi_1})\rangle+{}\nonumber\\&&
bc|\bar{A}(\bar{\theta_1},\bar{\phi_1})\rangle|A(\theta_1,\phi_1)\rangle+bd|\bar{A}(\bar{\theta_1},\bar{\phi_1})\rangle|\bar{A}(\bar{\theta_1},\bar{\phi_1})\rangle\longrightarrow{}\nonumber\\&&
ac|A(\theta_1,\phi_1)\rangle|A(\theta_1,\phi_1)\rangle+ad|A(\theta_1,\bar{\phi_1})\rangle|\bar{A}(\bar{\theta_1},\phi_1)\rangle+{}\nonumber\\&&
bc|\bar{A}(\bar{\theta_1},\phi_1)\rangle|A(\theta_1,\bar{\phi_1})\rangle+bd|\bar{A}(\bar{\theta_1},\bar{\phi_1})\rangle|\bar{A}(\bar{\theta_1},\bar{\phi_1})\rangle
\end{eqnarray}
\\
However if we consider the qubit $|X(\theta,\phi)\rangle$ and it's
orthogonal state $|\bar{X}(\bar{\theta},\bar{\phi})\rangle$ at the
input port, then the perfect swapping of phase angles between
these two states will be given by the transformation,
\begin{eqnarray}
&&|X(\theta,\phi)\rangle|\bar{X}(\bar{\theta},\bar{\phi})\rangle\longrightarrow|X(\theta,\bar{\phi})\rangle|\bar{X}(\bar{\theta},\phi)\rangle={}\nonumber\\&&
\{\alpha|A(\theta_1,\phi_1)\rangle+\beta|\bar{A}(\bar{\theta_1},\bar{\phi_1})\rangle
\}\{\gamma|A(\theta_1,\phi_1)\rangle+\delta|\bar{A}(\bar{\theta_1},\bar{\phi_1})\rangle\}={}\nonumber\\&&
\alpha\gamma|A(\theta_1,\phi_1)\rangle|A(\theta_1,\phi_1)\rangle+\alpha\delta|A(\theta_1,\phi_1)\rangle|\bar{A}(\bar{\theta_1},\bar{\phi_1})\rangle+{}\nonumber\\&&
\beta\gamma|\bar{A}(\bar{\theta_1},\bar{\phi_1})\rangle|A(\theta_1,\phi_1)\rangle+\beta\delta|\bar{A}(\bar{\theta_1},\bar{\phi_1})\rangle|\bar{A}(\bar{\theta_1},\bar{\phi_1})\rangle
\end{eqnarray}
where
\begin{eqnarray}
\alpha^2+|\beta|^2=1\nonumber\\
\gamma^2+|\delta|^2=1
\end{eqnarray}
It is clearly evident that the equations (6) and (7) are not
identical for all values of the two parameters with full
generality. This indicates that linear structure of quantum
mechanics doesn't allow swapping of the phase angles for unknown
orthogonal qubits.\\\\\\
Next we will see see that quite alike to the previous case one
cannot swap the azimuthal angles between two unknown orthogonal
qubits $|X(\theta,\phi)\rangle,
|\bar{X}(\bar{\theta},\bar{\phi})\rangle$. Let us once again
assume that the swapping of the azimuthal angles is possible for
orthogonal vectors $\{
|A(\theta_1,\phi_1)\rangle,|\bar{A}(\bar{\theta_1},\bar{\phi_1})\rangle\}$,
\begin{eqnarray}
|A(\theta_1,\phi_1)\rangle|A(\theta_1,\phi_1)\rangle\longrightarrow|A(\theta_1,\phi_1)\rangle|A(\theta_1,\phi_1)\rangle\nonumber\\
|A(\theta_1,\phi_1)\rangle|\bar{A}(\bar{\theta_1},\bar{\phi_1})\rangle\longrightarrow|A(\bar{\theta_1},\phi_1)\rangle|\bar{A}(\theta_1,\bar{\phi_1})\rangle\nonumber\\
|\bar{A}(\bar{\theta_1},\bar{\phi_1})\rangle|A(\theta_1,\phi_1)\rangle\longrightarrow|\bar{A}(\theta_1,\bar{\phi_1})\rangle|A(\bar{\theta_1},\phi_1)\rangle\nonumber\\
|\bar{A}(\bar{\theta_1},\bar{\phi_1})\rangle|\bar{A}(\bar{\theta_1},\bar{\phi_1})\rangle\longrightarrow|\bar{A}(\bar{\theta_1},\bar{\phi_1})\rangle|\bar{A}(\bar{\theta_1},\bar{\phi_1})\rangle
\end{eqnarray}
Now if we apply this swapping transformation defined for the
orthogonal basis vectors $\{
|A(\theta_1,\phi_1)\rangle,|\bar{A}(\bar{\theta_1},\bar{\phi_1})\rangle\}$,
on the product state
$|X(\theta,\phi)\rangle|\bar{X}(\bar{\theta},\bar{\phi})\rangle$,
the output is given by,
\begin{eqnarray}
&&|X(\theta,\phi)\rangle
|\bar{X}(\bar{\theta},\bar{\phi})\rangle={}\nonumber\\&&\{a|A(\theta_1,\phi_1)\rangle+b|\bar{A}(\bar{\theta_1},\bar{\phi_1})\rangle\}
\{
c|A(\theta_1,\phi_1)\rangle+d|\bar{A}(\bar{\theta_1},\bar{\phi_1})\rangle\}={}\nonumber\\&&
ac|A(\theta_1,\phi_1)\rangle|A(\theta_1,\phi_1)\rangle+ad|A(\theta_1,\phi_1)\rangle|\bar{A}(\bar{\theta_1},\bar{\phi_1})\rangle+{}\nonumber\\&&
bc|\bar{A}(\bar{\theta_1},\bar{\phi_1})\rangle|A(\theta_1,\phi_1)\rangle+bd|\bar{A}(\bar{\theta_1},\bar{\phi_1})\rangle|\bar{A}(\bar{\theta_1},\bar{\phi_1})\rangle\longrightarrow{}\nonumber\\&&
ac|A(\theta_1,\phi_1)\rangle|A(\theta_1,\phi_1)\rangle+ad|A(\bar{\theta_1},\phi_1)\rangle|\bar{A}(\theta_1,\bar{\phi_1})\rangle+{}\nonumber\\&&
bc|\bar{A}(\theta_1,\bar{\phi_1})\rangle|A(\bar{\theta_1},\phi_1)\rangle+bd|\bar{A}(\bar{\theta_1},\bar{\phi_1})\rangle|\bar{A}(\bar{\theta_1},\bar{\phi_1})\rangle
\end{eqnarray}
Now if we consider the swapping of the azimuthal angle of two
unknown orthogonal qubits
$|X(\theta,\phi)\rangle,|\bar{X}(\bar{\theta},\bar{\phi})\rangle$,
the final state after the swapping is given by,
\begin{eqnarray}
&&|X(\theta,\phi)\rangle|\bar{X}(\bar{\theta},\bar{\phi})\rangle\longrightarrow|X(\bar{\theta},\phi)\rangle|\bar{X}(\theta,\bar{\phi})\rangle={}\nonumber\\&&
\{\alpha_1|A(\theta_1,\phi_1)\rangle+\beta_1|\bar{A}(\bar{\theta_1},\bar{\phi_1})\rangle
\}\{\gamma_1|A(\theta_1,\phi_1)\rangle+\delta_1|\bar{A}(\bar{\theta_1},\bar{\phi_1})\rangle\}={}\nonumber\\&&
\alpha_1\gamma_1|A(\theta_1,\phi_1)\rangle|A(\theta_1,\phi_1)\rangle+\alpha_1\delta_1|A(\theta_1,\phi_1)\rangle|\bar{A}(\bar{\theta_1},\bar{\phi_1})\rangle+{}\nonumber\\&&
\beta_1\gamma_1|\bar{A}(\bar{\theta_1},\bar{\phi_1})\rangle|A(\theta_1,\phi_1)\rangle+\beta_1\delta_1|\bar{A}(\bar{\theta_1},\bar{\phi_1})\rangle|\bar{A}(\bar{\theta_1},\bar{\phi_1})\rangle
\end{eqnarray}
where
\begin{eqnarray}
\alpha_1^2+|\beta_1|^2=1\nonumber\\
\gamma_1^2+|\delta_1|^2=1
\end{eqnarray}
Again we observe that equations (10) and (11) are not identical
with full generality. It is clearly evident that we cannot swap
the azimuthal angles of two unknown orthogonal qubits.\\
The above proof only deals with qubits, however for a more general
structure one should consider qudits. Arbitrary quantum states
$|\psi(\theta,\phi)\rangle,|\bar{\psi}(\bar{\theta},\bar{\phi})\rangle\in
H$, (where $H$ is the d dimensional Hilbert space) can be
expressed in terms of the basis vectors
$\{|A_i(\theta_i,\phi_i)\rangle, i=1,....d\}$. Even if we assume
that we can partially swap the phase angles and the azimuthal
angles for the basis vectors
$\{|A_i(\theta_i,\phi_i)\rangle,|A_j(\theta_j,\phi_j)\rangle\}$,
then on the basis of the principle of linearity of quantum
mechanics it is not possible to partially swap the same parameters
 for arbitrary quantum states
$|\psi(\theta,\phi)\rangle,|\bar{\psi}(\bar{\theta},\bar{\phi})\rangle$.
This is because of the fact that the states obtained after partial
swapping of the parameters of the two states
$|\psi(\theta,\phi)\rangle,|\bar{\psi}(\bar{\theta},\bar{\phi})\rangle$
is not going to be identical with the output obtained if we
consider the linear combination of the two states in terms of
their basis vectors and apply partial swapping for the basis
vectors.
\section{Conclusion:}In summary we can say that here we have
introduced a new kind of impossible operation in quantum
information theory, which we refer as the partial swapping of
quantum parameter. It has already been observed that many quantum
information processing protocol which are Hilbert space or for the
equatorial states of the Bloch sphere , are not true for the
entire complex Hilbert space. These operations are like remote
state preparation [6], rotation of real angles in Grover's
algorithm [7] and many. This no go theorem tells us that one
cannot separately swap two parameters $\theta$ and $\phi$. This
theorem provides a strong evidence that all such tasks that are
possible in real Hilbert space can never be implemented on the
entire complex Hilbert space.
\section{Acknowledgement:} I.C acknowledges Prof B.S.Choudhury, Bengal Engineering and
Science University for providing encouragement in the completion
of this work. I.C also acknowledges Prof C.G.Chakraborti, S. N
Bose Professor of Theoretical Physics, University of Calcutta, for
being the source of inspiration in carrying out research. I.C also
acknowledges S. Adhikari for having useful discussions for the
improvement of the work.
\section{Reference:}
$[1]$ W.K.Wootters and W.H.Zurek,Nature \textbf{299},802(1982).\\
$[2]$ A.K.Pati and S.L.Braunstein, Nature \textbf{404},164(2000).\\
$[3]$ A.K.Pati and S.L.Braunstein, e-print quant-ph/0303124.\\
$[4]$ A.K.Pati and Barry C.Sanders,  Phys. Lett. A \textbf{359},
31-36 (2006).\\
$[5]$ Duanlu Zhou, Bei Zeng, and L. You, Phys. Lett. A \textbf{352}, 41 (2006).\\
$[6]$ A.K.Pati, Phys. Rev. A \textbf{63},014302 (2001).\\
$[7]$ L.K.Grover, Phys.Rev.Lett. 78, 325 (1997).
\end{document}